\documentclass[10pt,a4paper]{article}
\usepackage{preamble}

\title{On the stability of string-hole gas}

\author[a]{Denis Bitnaya,}
\author[a]{Pietro Conzinu,}
\author[a]{and Giovanni Marozzi}

\affiliation[a]{Dipartimento di Fisica, Universit\`a di Pisa, Largo B. Pontecorvo 3, 56127 Pisa, 
Italy,\\
and Istituto Nazionale di Fisica Nucleare, Sezione di Pisa, Largo B. Pontecorvo 3, 56127 Pisa, Italy.}

\emailAdd{d.bitnaya@studenti.unipi.it}
\emailAdd{pietro.conzinu@phd.unipi.it}
\emailAdd{giovanni.marozzi@unipi.it}

\abstract{ 
Focusing on a string-hole gas within the pre-big bang scenario, we study the stability of its solutions in the phase space. 
We firstly extend the analysis present in the literature relaxing the ideal-gas properties of the string-hole gas, taking into account a (bulk-)viscosity term. Then we consider the case of a theory described by a complete O(d,d)-invariant action up to all orders in 
$\ap$-corrections (the Hohm-Zwiebach action), studying the stability of the string-hole gas solution with or without the introduction of the viscosity term. 
Furthermore, the bulk viscosity is also considered for two different first order $\ap$-corrected actions: the Gasperini-Maggiore-Veneziano-action and the 
Meissner-action.
The results obtained show how the viscosity can help to stabilize  the string-hole gas solution, obtaining constraints on the equation of state of the gas.}
\makeatletter
\gdef\@fpheader{}
\makeatother
\begin{document}
\maketitle

\section{Introduction}
The construction of cosmological scenarios embedded in string theory is a main open issue within modern cosmology. To this purpose, the 
 use of symmetries, usually called dualites in string theory, is crucial~\cite{Brandenberger:1988aj,Veneziano:1991ek, Meissner:1991zj}.
 In particular, the T-duality~\cite{Brandenberger:1988aj, Veneziano:1991ek} states that given a universe with radius $R$, the considered theory is invariant under the transformation $R \rightarrow l^2_s/R$, where $l_s$ is the string length. This provides a minimum length ($R\sim l_s$) and eliminates the singularity issue, leading naturally to pre-big bang scenarios~\cite{Gasperini:2007vw, Brandenberger:2008nx, Gasperini:2021mat, Gasperini:2007zz,Brandenberger:2006vv}.
In a recent study~\cite{Quintin:2018loc}, it was proposed that the final state of the pre-big bang epoch is dominated by a black-hole gas in a quasi-static Hagedorn phase~\cite{Atick:1988si}. Such a gas of black (string)-hole could naturally form by instabilities in a contracting universe (see e.g.~\cite{Quintin:2016qro,Conzinu:2020cke,Conzinu:2023fui}). In particular, in~\cite{Quintin:2018loc} was described the stability of the black-hole gas solution in the presence of a dilaton potential and considered the first $\ap-$correction contribution for two different actions: the Gasperini-Maggiore-Veneziano (GMV)-action~\cite{Gasperini:1996fu} and the Meissner-action~\cite{Meissner:1996sa}, with $\ap=l^2_s/2 \pi$ the dimensionful parameter associated with the string length $l_s$ . These $\ap-$corrections account for the leading-order corrections to the classical equations of motion from the non-zero string length.
Therefore they are essential for a more accurate description of the pre-big bang phase, when the curvature reaches the string scale.

Following this seminal work, we are going here to extend the results present in~\cite{Quintin:2018loc} in different ways.
First, we will include $\ap-$corrections up to all orders, using the duality-invariant action presented in~\cite{Hohm:2019jgu} (see also~\cite{Bernardo:2019bkz}) to describe the string-hole gas matter. 
As mentioned, higher-order corrections are necessary when $H^{-1} \sim l_s$, in this case all corrections become comparable and one cannot truncate the expansion.
In~\cite{Hohm:2019jgu} Hohm and Zwiebach addressed this issue by assuming that the O(d,d)-duality holds to all orders, obtaining a simple and systematic action for these $\ap-$corrections. 
This duality-invariant action is a significant advancement in the context of string cosmology, allowing
for a more consistent and robust description of the cosmological evolution.
Furthermore, to enhance the stability of the solution, we will also consider a (bulk-)viscosity term to relax the ideal-gas properties of the string-hole gas. The introduction of a viscosity term in this framework is motivated by several factors. First of all, it can be seen
as an attempt to improve the description of the fluid itself. Considering a realistic description of the dynamic of a gas, the existence
of dissipative terms can be traced back to an interaction between the gas particles.
For example, in the context of a black-hole gas the authors of~\cite{Ganguly:2021pke} have explicitly shown how a viscosity terms can emerge from gravitational interaction.
Moreover, it is important to note that a perfect fluid description is not stable under general O(d,d) transformations~\cite{Gasperini:1991ak, Gasperini:2007zz}, indeed a general transformation could induce bulk/shear viscosity terms.
Here, we will focus only on the contribution of the bulk viscosity term, since we are considering spatial homogeneous and isotropic backgrounds 
(but see, for example,~\cite{Quintin:2021eup, Fanizza:2022pvx} for possible effects of shear viscosity in anisotropic scenarios). 
We leave the investigation of more general description with shear-viscosity for future analysis. This approximation should be appropriate for a preliminary analysis and serves as a basis for more comprehensive studies.

The manuscript is organized as follows. In Sect.~\ref{sec.2} we review the properties of a string-hole gas (SHG).
In Sect.~\ref{sec.3} we introduce the non-perfect fluid description and then we study the SHG solution in the presence of bulk viscosity at $0$-order in $\ap$-correction (low-energy action).
In Sect.~\ref{sec.4} we study the SHG-solution in a complete O(d,d)-invariant action up to all orders in $\ap$, with and without a bulk viscosity term. 
Finally, in Sec.~\ref{sec.5} we study the impact of a bulk viscosity term at first-order in $\ap-$corrections for the GMV-action and for the Meissner-action. 
In Sect.~\ref{sec.6} we present our final remarks and conclusions, while in 
Appendix~\ref{Ap.A} we give some useful relations.

\section{Definition of a String-Hole gas}
\label{sec.2}
We recall that a string-hole~\cite{Damour:1999aw,Veneziano:2012yj} is a Schwarzschild black-hole~\cite{Emparan:2008eg}
confined within a radius given by the string length ($R_{SH} = \Ls$), such that its mass is 
\begin{equation}
    M_{SH}\sim \frac{R_{SH}^{D-3}}{G} \sim \frac{\Ls^{D-3}}{G}\,,
\end{equation}
with $D$ the total space-time dimension.
By taking in consideration the following tree level relation~\cite{Gasperini:2007zz} 
 \begin{equation}
 \left(\frac{l_\mrm{Pl}}{l_\mathrm{s}}\right)^{D-2}=\left(\frac{M_\mathrm{s}}{M_\mathrm{Pl}}\right)^{D-2}=e^\phi=g_\mrm{s}^2\ll 1~,
\end{equation}  
where $\f$ is the dilaton field, $ \Ms =l_\mrm{s}^{-1} $ the string mass and $M_\mathrm{Pl}^{2-D}=8\pi G$ the Planck mass, one can easily see that the string-hole mass can be rewritten as
\begin{equation}
    M_{SH} \sim \Ms \gs^{-2} \,,
\end{equation} 
 which is the so-called correspondence curve~\cite{Veneziano:2003sz}, where strings and black-holes share remarkable similarities (see e.g.~\cite{Veneziano:2003sz,Quintin:2018loc}). 
Therefore, the appropriate description of an universe populated with a dense gas of black-holes at the string scale must be a string-hole gas (SHG).

Let us now consider a dense gas composed by N string-holes with single energy $E_\mrm{SH}=M_\mrm{SH}$ and entropy $S_\mrm{SH}=l_\mrm{s} E_\mrm{SH}$~\cite{Quintin:2018loc}. 
The energy and entropy of this SHG will then be provided by
\begin{equation}
E_{gas}=E=NE_\mrm{SH} \sim N \Ls^{-1}e^{-\phi}\, ,\qquad   S_{gas}=S=NS_\mrm{SH} \sim Ne^{-\phi} \,.
\end{equation}
The physical volume of the gas can be written as
 \begin{equation}
   V_{gas}\equiv V=\g N V_\mrm{SH}\,,
 \end{equation}
 where $\g$ quantify the ``density'' of the gas, and $V_\mrm{SH} \sim \Ls^{D-1} $ is the volume of a single string-hole.
 Thus the number $N$ is given by $N \sim V l_s^{1-D}$, such that the energy and entropy densities are~\footnote{Here we have used the following relation: $\ e^{\phi} \sim G \Ls^{2-D}$.}

\begin{equation}
    \rho={E \over V} \sim \Ls^{-D}e^{-\phi} \sim \Ls^{-2} G^{-1}\,,  \qquad 
    s= {S\over V} \sim \Ls^{1-D}e^{-\phi}\sim \Ls^{-1}G^{-1} \sim \Ls \rho\,\,.
\end{equation}

On the other hand, the generalized second law of thermodynamics in the presence of dilaton charge $\s$ reads~\footnote{ This expression, that we write in terms of O(d,d) invariant variables, came from the request that one should obtain for $dS=0$  the 
 standard continuity equation. On the other hand, for zero dilaton charge and in the presence of viscosity, one should be able to obtain the standard result for entropy production (following, for example, the Weinberg argument of \cite{Weinberg:1971mx}), as we will show in the next section.}
\begin{equation}\label{eq 1principio}
    TdS=dE+ \bar{p}_I dV -\frac{\bar \sigma}{2} d \Phi \,,
\end{equation}
where $ \Phi=\phi-d \log a$ is the shifted dilaton, 
a quantity $\bar{A}$ is defined as $\bar{A}\equiv a^d A$ and we also define $p_I\equiv p-\frac{\sigma}{2}$ \footnote{$p_I$ is the O(d,d)-invariant pressure (see \cite{Quintin:2021eup} for a more in deep discussion).}. As a consequence one obtains that
\begin{equation}
    p_I=T \frac{\partial S}{\partial V}\Big|_{\Phi, E}  \, , \qquad \qquad  \frac{\bar \sigma}{2}=-T \frac{\partial S}{\partial \Phi}\Big|_{V,E}\,. 
\end{equation}
Furthermore, using the relation $\frac{1}{T}= (\frac{\partial S}{\partial E})_{V,\Phi}$ one finds that the temperature is proportional to the Hagedorn temperature $
 T_{Hag} \sim \Ls^{-1}  \,
$. 

Considering a FLRW-universe with scale factor $a$, and requiring $\rho\sim a^{-d}$, we have that $a\sim G^{\frac{1}{d}} \sim e^{\frac{\phi}{d}} $ and obtain the following Hubble factor:
\begin{equation}\label{2.9}
    H=\frac{\dot{a}}{a}=\frac{\dot{\phi}}{d}\,,
\end{equation}
that equivalently can be expressed as $\dot {\Phi}=0$. On the other hand, using the continuity equation
\begin{equation}\label{eq cont}
    \dot{\bar\rho}+dH \bar{p}_I=\frac{\bar \sigma}{2} \dot {\Phi} \,,
\end{equation} 
one then obtains
\beq
p_I=0\,,
\eeq
which gives us the relation $\s=2p$\,.

Therefore, if the string-hole radius $R_{SH}\sim l_S$ is proportional to the Hubble radius $R_{SH}\propto H^{-1}$, 
the evolution of a string-hole gas in the string frame corresponds to a constant Hubble parameter equal to the string mass and a linearly growing dilaton field~\cite{Quintin:2018loc}.

\section{Non-perfect string-hole fluid}
\label{sec.3}
Due to gravitational attraction a gas of black-hole can be seen as a viscous fluid (see, for example,~\cite{Ganguly:2021pke}). 
Therefore,
we go now beyond the perfect fluid approximation, adding the viscosity to the model considered.

To begin we briefly review the description of a generic fluid in presence of viscosity (we follow the presentation of~\cite{Ganguly:2021pke}).
Given a time-like fluid
4-velocity $u^a$, such that $g_{ab}u^au^b=-1 $, one can define the correspondent induced metric on the spatial hypersurface by
\[ h_{ab}=g_{ab} + u_au_b\,,\]
and the extrinsic curvature by
\[  K_{ab}= h_a^ch_b^d\nabla_du_c=D_bu_a=\nabla_b u_a+u_b\dot u_a \,,\]
where $\dot u_a=u^c \nabla_c u_a$. 
The extrinsic curvature can be then decomposed in its symmetric and antisymmetric part as
\[ 
K_{ab}=\Theta_{ab}+\omega_{ab}  \,,
\]
where the symmetric part $\Theta_{ab} $ is called the expansion tensor, while the antisymmetric part $\omega_{ab} $ is called the vorticity tensor.
For simplicity we will assume in the follow a vanishing antisymmetric part $\omega_{ab}=0$. 
We can then further expand $\Theta_{a b}$ 
in the following
irreducible representations:
\beq
 \Theta_{ab}=\frac{1}{2}\Theta h_{ab}+\sigma_{ab}\,,
 \eeq
with $\Theta $ the trace part of the expansion tensor
and $\sigma_{ab} $ its traceless part called shear tensor, i.e.
\beq
\Theta=g^{ab}\Theta_{ab}=D_au^a=\nabla_a u^a \qquad,\qquad g^{ab}\sigma_{ab}=0\,.
\eeq

The energy-momentum tensor for a generic fluid can then be written in terms of this expansion tensor (neglecting heat transfer) as~\cite{Quintin:2018loc}
\beq
    T_{ab}=\rho\, u_a u_b+(p-\zeta \Theta )h_{ab}-2\eta \sigma_{ab}\,,
\label{TmunuWithViscosity}
\end{equation}
 where the coefficients $\eta$ and $\zeta$ are respectively the shear viscosity and the bulk viscosity.
Here we are not interested in anisotropies and so, for simplicity, we will restrict our analysis to a zero shear viscosity.
Considering only the bulk viscosity
we can then define an effective pressure $p_\mrm{eff}$ 
in the following way
\begin{equation}
\label{general effective pressure}
    p_\mrm{eff}= p-\zeta \Theta\,.
\eeq

 Our purpose is to describe a SHG (in the S-frame), 
 including the presence of a bulk viscosity $\zeta$, as a fluid with effective pressure $p_\mrm{eff}=\s/2$, without constraints on the equation of state $p=\om \rho$.
This is realized, in a homogeneous FLRW background (where $\Theta = d H $), still assuming $\r \sim a^{-d}$ such that $ \dot {\Phi}=0$ (see Eq.~\eqref{2.9}). In fact, using the continuity equation
\begin{equation}
    \dot{\bar\rho}+dH (\bar{p}_I-d H \bar \z)=\frac{\bar \sigma}{2} \dot {\Phi} \,,
\end{equation} 
we obtain
\beq
    p_I-d H \z=0\,,
\eeq
which gives $p_\mrm{eff}=\sigma/2$.

Note that, as we can directly verify combining Eq.~\eqref{eq 1principio} and Eq.~\eqref{eq cont}, in the presence of bulk viscosity it holds (see~\cite{Weinberg:1971mx} for similar conclusions)
\begin{equation}
    T \dot S =d^2H^2 \zeta \,,
\end{equation}
such that we are in a period of entropy production. 
Therefore, during the evolution it is expected that at some point, when we saturate the entropy bounds,  $\z \rightarrow 0 $. 
For our purposes it is enough to consider a particular regime of constant viscosity, which can be seen as a good approximation assuming a slow evolution of the viscosity term. In this way we describe, de facto, only a first period of the full evolution.
Furthermore, in our description of this first period of the full evolution we also assume a negligible variation of the dilaton field $\phi$. Namely, we assume that $e^{\phi} \zeta$ is roughly constant.

\subsection{Low-energy action}
Let us
start our analysis applying the non-perfect fluid description (considering only a bulk viscosity term) to the low-energy action and, following~\cite{Quintin:2018loc}, considering the contribution of a generic dilaton potential
\beq
\label{EQ.3.8}
S=-\frac{1}{2\Ls^{d-1}}\int d^{d+1}x\sqrt{|g|} e^{-\phi}(R+g^{\mu\nu}\nabla_{\mu} \phi \nabla_{\nu} \phi +2\Ls^{d-1}U(\phi) ) +S_m \,,  
\eeq
where we explicitly use $D=d+1$, with $d$ the space dimensions. 
It should be noted that the matter action $S_m$ in Eq.~\eqref{EQ.3.8} is just a formal action, since hydrodynamics, especially in the presence of dissipative terms, it is usually formulated directly in terms of equations of motion instead of an action principle (see e.g.~\cite{Son:2007vk}).

The main effect of the bulk viscosity $\z$ is to modify the pressure, such that the equations of motion for a FLRW-metric become the following
\begin{eqsplit}
    d(d-1)H^2+\dot{\phi}^2-2dH\dot {\phi}=&2\Ls^{d-1}(e^{\phi}\rho +U(\phi))\,,\\
    \dot H-H\dot \phi +dH^2=&\Ls^{d-1}\le(e^{\phi} \le(p_{eff}-\frac{\sigma}{2}\ri)-U_{,\phi}  \ri) \,,\\
    2\ddot{\phi}-\dot{\phi}^2+2dH\dot{\phi}-2d\dot H -d(d+1)H^2=&2\Ls^{d-1}(e^{\phi}\frac{\sigma}{2}-U(\phi)+U_{,\phi} )\,.
\end{eqsplit}
Assuming a barotropic equation of state $ p= \omega \rho $ and recalling the string-hole solution
\beq \label{SH-solution}  
\rho=\rho_0 a^{-d}= C \Ls^{-d-1}e^{-\phi}\,,\qquad H=\frac{\dot \phi}{d}=\mrm{const.}\,, \qquad  \sigma=2 p_{eff}\,,
\eeq
 where $C$ is a constant, the equations of motion become
\begin{eqsplit}\label{4.6}
    -dH^2&=2\Ls^{d-1}(e^{\phi}\rho+U) \,, \\
    0&=\Ls^{d-1} U_{,\phi} \,, \\
    -dH^2&=2\Ls^{d-1}(e^{\phi}p_{eff}-U)\,.
\end{eqsplit}

Assuming that $H^{-1} $ is of order of the string length, from the first equation of Eq.~\eqref{4.6}, 
the potential $U$ is fixed to be
\begin{equation}
    U\sim -\Ls^{-d-1}\le[C+\frac{d}{2}\ri]\,, 
\end{equation}
which is equivalent to the addition of a negative cosmological constant $\Lambda \sim \mathcal{O}(\Ls^{-d-1} ) $ to the problem, as it was already seen in~\cite{Quintin:2018loc} without a bulk viscosity term.
As a consequence, the low energy action seems not sufficient to describe a SHG, at least without adding a fine-tuned negative cosmological constant in the potential.
Therefore, hereafter we will extend this scenario taking into account also $\ap-$corrections.

\section{Complete O(d,d)-invariant action to all orders in $\ap$-corrections with and without viscosity}
\label{sec.4}

In the follow we extend the results obtained in~\cite{Quintin:2018loc} to the case of an action valid to all order in $\ap$ (see e.g.~\cite{Bernardo:2019bkz,Hohm:2019jgu,Quintin:2021eup,Bernardo:2020zlc}), studying the stability of the SHG solution. 
As mentioned, there are several motivations for considering $\ap-$corrections to all orders. 
On one hand,
 when we approach string-curvature scales $H^{-1} \sim l_S$, all higher-order corrections in $\ap$ become comparable to each other. On the other hand, if we try to truncate the series in $\ap$, for example at first order, 
 one can find a large amount of
actions related by a simple fields redefinition~\cite{Gasperini:2007zz}, with the consequent ambiguity regarding the right choice to adopt. Fortunately, such ambiguity vanishes when we include all the $\ap$ corrections.
 Hohm and Zwiebach adressed this issue by assuming that the O(d,d)-duality~\cite{Meissner:1991zj} holds to all orders, obtaining a simple and systematic action for the $\ap-$corrections~\cite{Hohm:2019ccp,Hohm:2019jgu}. This duality-invariant action is a significant advancement in the context of string cosmology, allowing
for a more consistent and robust description of the cosmological evolution
\cite{Hohm:2019jgu,Hohm:2019ccp,Bernardo:2019bkz,Quintin:2021eup,Bernardo:2020zlc,Codina:2020kvj,Codina:2021cxh,Wang:2019kez,Wang:2019dcj,Codina:2023fhy,Nunez:2020hxx, Gasperini:2023tus, Conzinu:2023fth}.

Let us briefly recall the notation we will use. The massless Neveu-Schwarz sector of all superstring theories is composed by 3 fields: the symmetric spacetime metric tensor $g_{\mu\nu}$, the antisymmmetric Kalb-Ramond tensorial field  $b_{\mu\nu}$ and the dilaton field $\phi$. For cosmological purpose we consider a purely time-dependent (d+1)-dimensional string background
\beq
g_{\mu\nu}= \begin{pmatrix}
-n^2(t) & 0\\
0 & g_{ij}(t)
\end{pmatrix}\,,\qquad \qquad b_{\mu\nu}= \begin{pmatrix}
0 & 0\\
0 & b_{ij}(t)
\end{pmatrix} \, ,\qquad \qquad \f= \f(t). 
\eeq
We consider then the following string action, up to all orders in $\ap$-corrections, coupled to matter in a O(d,d)-invariant way (see e.g.~\cite{Quintin:2021eup,Bernardo:2020zlc})~\footnote{Let us underline that this is just an assumption, but the SHG could in principle not be invariant under such a symmetry.}
\begin{equation}\label{Eq3.2}
    S[\Phi, n, S, \chi]=\frac{1}{2\kappa^2}\int d^d x dt n e^{-\Phi}\le[-(\mathcal D_t \Phi)^2+\sum_{k=1}^{\infty}(\ap)^{k-1}c_k \text{tr}\le((\mathcal D_t\mc S )^{2k}\ri)\ri]+S_m[\Phi, n, S,\chi ]\,,
\end{equation}
where $ \chi$ is a generic matter field, $ \kappa \propto l_s^{d-1} $ defines the 
D-dimensional Newton constant, $\mc D_t\equiv 1/n \, \pa_t$, $\Phi$ is the shifted dilaton defined as 
\begin{equation}
    \Phi=\phi-\log\sqrt{g_s}\,,
\end{equation}
and $\mc S$ is an O(d,d) invariant $2 d \times 2 d$ matrix given by~\footnote{Note that $g_s\equiv det (g_{i j})$, $g\equiv det (g_{\mu\nu})$ and we have $\sqrt{-g} =n \sqrt{g_s}$.}
\begin{equation}
    \mc S=
    \begin{pmatrix}
    & bg^{-1} & g-bg^{-1}b \\
    & g^{-1} & -g^{-1}b \\
    \end{pmatrix}\,,
\end{equation}
namely this is constructed by the two-form field  $b_{ij} $, the spatial metric $g_{ij} $ and its inverse $g^{ij} $,  such that $\mc S^2=1$.
We can always define the energy-momentum tensor and
charge density of the scalar field as
\begin{equation}
    T_{\mu\nu}=-\frac{2}{\sqrt{-g}}\frac{\delta S_m}{\delta g^{\mu\nu}}\,, \qquad \qquad 
    \sigma=-\frac{2}{\sqrt{-g}}\le(\frac{\delta S_m}{\delta \Phi}\ri)\,,
\end{equation}
and we consider a generic isotropic fluid, with eventually a bulk viscosity term, such that the pressure $p$ can in general contains the viscosity term as in Eq.~\eqref{general effective pressure}.

Let us assume, for simplicity, the case of a flat FLRW background \[ n(t)=1\,, \qquad   g_{ij}(t)=a(t)^2\d_{ij}\, \qquad b_{ij}=0\,, \]
such that the equations of motion take the following form (see e.g.~\cite{Bernardo:2019bkz})
\begin{eqsplit}
    \dot {\Phi}^2+H F'(H)-F(H)=& Y \bar \rho \,,\\
    \dot H F''(H)-\dot \Phi F'(H)=&-Yd\le(\bar{p}-\frac{\bar{\sigma}}{2} \ri) \,, \\
    2\ddot \Phi -\dot \Phi^2 +F(H)=& Y \frac{\bar{\sigma }}{2} \,, 
\end{eqsplit}
while the continuity equation is given by~\footnote{Naturally, these equations are invariant under the duality transformation $a \rightarrow \frac{1}{a} $ 
\begin{equation}
    H \rightarrow -H \qquad \Phi \rightarrow \Phi \qquad F(H) \rightarrow F(H) ,\qquad \bar \rho \rightarrow \bar \rho ,\qquad  \bar p \rightarrow -\bar p ,\qquad \bar \sigma \rightarrow \bar \sigma.
\end{equation}}
\begin{equation}
    \dot {\bar{\rho} } +dH \bar{p}_{I} -\frac{1}{2}\dot \Phi \bar \sigma=0\, ,
\end{equation}
where we define $Y\equiv 2\kappa^2 e^{ \Phi} $, and the function $F(H)$ as~\footnote{The coefficient $c_1= -\frac{1}{8} $ is fixed by the low energy description while the higher coefficients
are only partially known~\cite{Hohm:2019jgu,Hohm:2019ccp,Bernardo:2019bkz}.}
\begin{equation}\label{coeff ck}
    F(H)=2d \sum_{k=1}^{\infty} (-\ap)^{k-1}c_k 2^{2k}H^{2k} \, ,
\end{equation}
with $(\cdots)^\prime$ that means derivation w.r.t. 
$H$.

Imposing the SHG solution $\dot \Phi=0= \dot \phi-dH$ 
the above equations become
\begin{eqsplit}
 &   HF'-F 
    =Y \bar{\rho}\,,\\
&    0=
    -Yd(\bar{p}-\frac{\bar{\sigma}}{2}) \, \, \Rightarrow \bar{p}=\frac{\bar{\sigma}}{2}  \,,\\
  &  F=
    Y \frac{\bar{\sigma}}{2}  \,,\\
\end{eqsplit}
and we finally obtain the two conditions
\begin{eqsplit}\label{sol all ord}
F=&Y \bar{p} \,,\\
F' =&\frac{Y}{H}(\bar{\rho}+\bar{p}) \,.
\end{eqsplit}
In particular, assuming a barotropic relation $p=\om \r$, considering the case $ \omega=0 $ and in absence of viscosity ($\z=0$), we obtain from the first of 
Eqs.~\eqref{sol all ord} that
\begin{equation}
    \bar {p}=\bar {\sigma\over 2}=0\quad\implies \quad F=0 \, ,
\end{equation} 
while from the second of Eqs.~\eqref{sol all ord} it holds
\begin{equation}
    F'=\frac{Y\bar \rho}{H}\,.
\end{equation}
It follows a direct constraint on $F$ and on the first derivative $F'$ , with the last one completely fixed by the density of the gas. This seem to confirm the fact that the function $F(H)$ should be
non-analytic in the complex-H plane, as argued in \cite{Gasperini:2023tus} (see also \cite{Conzinu:2023fth}).

On the other hand, for $\om=0 $ but with a non zero bulk viscosity, we obtain $ \bar \sigma= -2 d \bar \zeta H $ such that 
\begin{eqsplit}
    F=&-Y d\bar \zeta H \, , \\
    F'=&\frac{Y}{H}(\bar \rho-d\bar \zeta H ) \, .
\end{eqsplit}
As a consequence, also in this case, we find a direct relation among $F(H)$, its first derivative and the density and viscosity of the gas.  

\subsection{Stability analysis of the fixed point}
Here, our purpose is to investigate the stability of the previously obtained string-hole gas solution. Specifically, we aim to establish whether this solution behaves as an attractor within the phase space of the theory.
To this aim, we firstly rewrite the equations of motion as: 
\begin{eqsplit}
     F\pr=&\frac{1}{H}[Y\bar{\rho}+F-\dot \Phi ^2 ]\,,  \\
    \dot H=& \frac{-Y d (\bar{p}-\frac{\bar{\sigma}}{2}) +F\pr\dot \Phi }{F''} \,, \\
    \ddot \Phi=&\frac{Y \frac{\bar{\sigma}}{2}+\dot \Phi^2-F}{2} \,,
\end{eqsplit}
where we are again using $p$ as a generic pressure that eventually can contain the viscosity contribution. 
Let us note that we can eliminate $F\pr$ using the first equation, such that the reduced system reads
\begin{eqsplit}
\dot H=& \frac{1}{F''}\left[-Yd\left(\bar p -\frac{\bar \sigma}{2}\right)+\frac{\dot \Phi}{H}( Y \bar \rho+F-\dot \Phi^2) \right] \, , \\
\ddot \Phi=& \frac{Y \frac{\bar\sigma}{2}+\dot \Phi^2-F}{2} \,. \\
\end{eqsplit}
Finally, imposing the condition $ \sigma=2 p $ one obtains
\begin{eqsplit}
\dot H=& \frac{1}{F''}\left[\frac{\dot \Phi}{H}( Y \bar \rho+F-\dot \Phi^2)\right] \, ,\\
\ddot \Phi=& \frac{Y \bar p+\dot \Phi^2-F}{2} \, .\\
\end{eqsplit}
The Jacobian of the above system is thus given by
\beq
\mathcal{J}=
\begin{pmatrix}
    \pa_H \dot H && \partial_{\dot \Phi} \dot H \\
    \partial_H \ddot \Phi && \partial_{\dot \Phi}\ddot \Phi 
\end{pmatrix}\,,
\eeq
with
\begin{subequations}
\begin{align}
\partial_H \dot H=&\frac{1}{F''^2} \left\{ \left[-\frac{\dot \Phi}{H^2}\left(Y \bar \rho +F-\dot \Phi^2 \right) +\frac{\dot \Phi}{H}F'\right]F''-\frac{\dot \Phi}{H}\left(Y \bar \rho +F -\dot{\Phi}^2 \right)F''' \right\}\,,\\
\partial_H \ddot \Phi=&\frac{Y \bar p ' -F'}{2}  \,,\\
\partial_{\dot \Phi}\ddot \Phi=& \dot \Phi \,,\\
\partial_{\dot \Phi} \dot H=&\frac{1}{HF''}[Y \bar \rho +F-3 \dot \Phi^2 ]\,.
\end{align}
\end{subequations}
where in the above derivatives we have considered two important points. First, that $Y \bar \rho $ is constant at the fixed point, and we can then neglect its derivative assuming to have a continuous and differentiable function at every point. Second, that we are considering a model for which $e^{\phi} \zeta$ can be considered constant (see Sect.~\ref{sec.3}), this translate in having $Y \bar{\zeta}$ equal to a constant and so independent from $H$ and $\dot{\Phi}$.
Now, imposing the solution in Eq.~\eqref{sol all ord} for $ F $ and $ F' $,  it follows that for any $\bar{p}$
\begin{eqsplit}
\partial_H \dot H=&0 \,,  \\
\partial_H \ddot \Phi= &-\frac{Y}{2 H}\bar \rho (1+ \omega) \,, \\
\partial_{\dot \Phi} \dot H=&\frac{Y}{HF''}[\bar \rho + \bar p ]\,, \\
\partial_{\dot {\Phi}} \ddot \Phi=&0\,,
\end{eqsplit}
and the trace and determinant of the Jacobian are the following 
\begin{equation}
tr \mc{J}=\partial_H \dot H +\partial_{\dot \phi}  \ddot \Phi=0 \,,
\label{TraceJ}
\end{equation}
\begin{equation}
det \mc{J}=\partial_H \dot H \partial_{\dot \Phi} \ddot \Phi-\partial_{\dot \Phi} \dot H \partial_H \ddot \Phi =-\frac{Y^2}{2H^2F''}(\bar \rho+\bar p) (1 +\omega) \bar \rho \,.
\label{DetJ}
\end{equation}
We can note that the trace is independent of $\bar{p}$, while the determinant when $\zeta=0$ reduce to
\begin{equation}
det \mc{J}= -\frac{Y^2}{2H^2F''}[\bar \rho(1+\omega)]^2 \,.
\end{equation}
Thus we can study the eigenvalues of the system distinguishing the cases with and without the viscosity contribution.
\begin{itemize}
    \item  {\bf Case with no viscosity:} for $\zeta=0$,
    $ det \mc{J} >0 $ implies $ F''<0 $  . 
Therefore, we have a null trace and a positive determinant under the  constraint $ F''<0 $. We can also rewrite these conditions in terms of the two eigenvalues $\l_1,\,\l_2$ of the jacobian $\mc{J}$, as 
\beq
\begin{cases}
\lambda_1+\lambda_2=0 \quad \rightarrow \quad \lambda_1=-\lambda_2 \\ 
    \lambda_1\cdot \lambda_2= -\frac{Y^2}{2H^2 F''}[\bar \rho(1+\omega)]^2>0
\end{cases} \,,
\eeq
such that the eigenvalues are
\beq
 \l_{1,2}= \pm i\frac{Y}{H\sqrt{2|F''|}}\bar{\rho}(1+\omega)\,.
\eeq

\item {\bf Case with viscosity:} on the other hand, when $ \zeta \ne 0 $, 
the eigenvalues $\lambda_1$ and 
$\lambda_2$ are given by
\begin{equation}
 \lambda_{1,2}= \pm i Y \frac{ \sqrt{(\omega +1)\bar{\rho }}   \sqrt{d H
   \bar{\zeta }-(\omega +1) \bar{\rho }}}{ H \sqrt{2 F''}}\,,
\end{equation}
and the stability requires that
\begin{equation}
det \mc{J}= -\frac{Y^2}{2H^2F'' } \bar \rho (1+\omega)[\bar \rho(1+\omega)-d \bar \zeta H]>0\,.
\label{detwV}
\end{equation}

Therefore, assuming $H$ and $\zeta$ both positives, Eq.(\ref{detwV}) is satisfied in the following two cases:
\begin{equation}
    -1<\omega<-1+\frac{d\zeta H}{\rho} \, , \qquad \qquad\text{if}\qquad F''>0 \,,
\end{equation}  

\begin{equation}
    \omega<-1 \, \vee \,  \omega>-1+\frac{d \zeta H}{\rho} \,, \qquad\text{if}\qquad F''< 0 \,.
\end{equation}
\end{itemize}
In the particular case $ \omega=0$ and $\zeta \ne 0  $, we have 
\begin{equation}
det J= -\frac{Y^2}{2H^2F'' } \bar \rho [\bar \rho-d \bar \zeta H]>0\,,
\end{equation}
which implies that
\begin{equation}
    \bar \rho< d \bar \zeta H \, ,  \qquad\text{if}\qquad F''> 0\,, 
\end{equation}
\begin{equation}
    \bar \rho> d \bar \zeta H \, ,   \qquad\text{if}\qquad F''< 0 \,.
\end{equation}

In summary, if we consider the SHG solution at all orders in $\ap$, with the Hohm-Zwiebach action, we find that the general form of the function $F(H)$ and its first derivative are constrained at the fixed point. 
In this case we have a ``center'' of the trajectory, since the real part of both eigenvalues is zero, we have circular orbits in the phase space, without touching the center of the circle. 
Therefore, the fixed point is not a real attractor, but a somekind of ``general'' stability is anyway achieved. To conclude, for $\zeta=0$ we have to require $F''< 0$, independently from the equation of state, while for $\z \neq 0$ we have more freedom, depending on the particular equation of state. \\

\section{Gasperini-Maggiore-Veneziano and Meissner actions: solution and stability}
\label{sec.5}
Hereafter, following~\cite{Quintin:2018loc}, we study the evolution of the SHG at first order in $\ap$. As said before, at first order in $\ap$ we have an ambiguity in the choice of the right action.
Here we will consider two particular cases, the Gasperini-Maggione-Veneziono (GMV) first order corrected-action  ~\cite{Gasperini:1996fu} 
(see also~\cite{Brustein:1997cv,Cartier:1999vk, Gasperini:2007zz, Quintin:2018loc}), given by
\begin{equation}
 S_{\alpha'} = \frac{k\alpha'}{8\ell_\mathrm{s}^{d-1}}\int\mathrm{d}^{d+1}x\,\sqrt{|g|}e^{-\phi}\left(\mathcal{G}-(\nabla_\mu\phi\nabla^\mu\phi)^2\right)~,
 \label{GMVaction}
\end{equation}
where $\mathcal{G}\equiv R_{\mu\nu\kappa\lambda}R^{\mu\nu\kappa\lambda}-4R_{\mu\nu}R^{\mu\nu}+R^2$
is the Gauss-Bonnet invariant. 
And the following manifestly O(d,d)-invariant action (from now on Meissner-action), firstly introduced by Meissner~\cite{Meissner:1996sa},
\beq
 S_{\alpha'} = ~\frac{k\alpha'}{8\ell_\mathrm{s}^{d-1}}\int\mathrm{d}^{d+1}x\,\sqrt{|g|}e^{-\phi}\Big(\mathcal{G}
  -(\nabla_\mu\phi\nabla^\mu\phi)^2 
-4G^{\mu\nu}\nabla_\mu\phi\nabla_\nu\phi+2(\nabla_\mu\phi\nabla^\mu\phi)\Box\phi\Big)~,
\label{Meissneraction}
\eeq
where $G_{\mu\nu}\equiv R_{\mu\nu}-Rg_{\mu\nu}/2$ is  the Einstein tensor 
and we use the d'Alembertian $\Box\equiv g^{\mu\nu}\nabla_\mu\nabla_\nu$ .

Those two actions are related by a field redefinition and, while the Meissner-action preserves the O(d,d) invariance, the GMV-action no. This is a direct evidence that in order to preserve this duality in an univocally way, it is necessary to take into account all the orders in $\ap$.  

We then couple those actions to a SHG fluid, with eventually a bulk viscosity term. 
Having that the equations of motion for both actions are the same in form~\cite{Gasperini:2007zz, Quintin:2018loc}, we firstly study the solution in a general way.

The equations of motion are
\bea\label{EOM first order O(d,d)}
\rho &&= C\Ls^{-1-d}e^{-\phi}=\frac{1}{2}\Ls^{1-d}e^{-\phi}\le(\dot \phi^2+d(d-1)H^2-2dH\dot\phi-\frac{3}{4}k\ap\mathcal{F}_{\rho}(H, \dot{\phi})\ri)\,,  \nonumber\\
\sigma &&=-\Ls^{1-d}e^{-\phi}\le(-2\ddot \phi +2d \dot H +\dot \phi^2 +d(d+1)H^2-2dH\dot \phi+\frac{k\ap}{4}\mathcal{F}_{\sigma}(H, \dot\phi, \dot H, \ddot \phi)\ri)\,,\nonumber\\
 p_\mrm{eff}&&=\frac{1}{2d}\Ls^{1-d}e^{-\phi}\le(-2d(d-1)\dot H+2d\ddot \phi-d^2(d-1)H^2+2d(d-1)H\dot \phi -d\dot \phi^2+\frac{k\ap}{4}\mathcal {F}_p(H,\dot \phi, \dot H, \ddot \phi)\ri)\,, \nonumber\\
 &&
\eea
where $p_\mrm{eff}$ is the effective pressure defined in Eq.~\eqref{general effective pressure}.
The functions $\mc{F}_\r\,, \, \mc{F}_\s\,,\, \mc{F}_p$, which are a combination of $\fdd$, $\fd$, $\Hd$ and $H$ which depends on the particular action, are defined in Appendix~\ref{Ap.A}.

Imposing the constraint given by the string-hole solution of Eq.~\eqref{SH-solution}, i.e. 
$\dot \phi=dH=\text{const.}$ and $\s=2 p_{eff}$,  a simpler system of equations follows as
\bea\label{5.4}
    \r=&&-{\tilde{Y}\over 2}\le[dH^2+\frac{3}{4}k\ap \mathcal F_{\rho} \ri]  \,,\nonumber\\
   \s= 2 p_\mrm{eff}=&&-\tilde{Y}\le[ d H^2+\frac{k\ap}{4 }\mathcal{F}_\s\ri]\,,\nonumber\\
    p_\mrm{eff}=&&-{\tilde{Y}\over 2} \le[dH^2-\frac{k\ap}{4 d}\mathcal{F}_{p}\ri] \,, 
\eea
where we define $\tilde{Y}\equiv e^{-\phi}\Ls^{1-d}$. Recalling that from Eqs.~\eqref{delta} one has 
$ \mathcal{F}_{\sigma}= \mathcal{F}_{\rho} $ and $ \mathcal{F}_{\rho}= -d H^4 \Delta $, we obtain after some manipulations the following two equations
\begin{subequations}
    \begin{align} 
 2\rho=2C\tilde{Y}=&\tilde{Y}(-dH^2+d\frac{3}{4}k\alpha' \Delta H^4)\,,\label{5.5a}\\
2\omega \rho-6\zeta H=&-\tilde{Y}(dH^2-k \frac{\alpha'}{4}d \Delta H^4 )\label{5.5b}\,.
\end{align}
\end{subequations}
By solving Eq.~\eqref{5.5a} for $H$ in terms of $C$ it follows
\beq \label{H generic}
 H=  \sqrt{ \frac{d+ \sqrt{d^2+8Cd\frac{3}{4}k \alpha' \Delta }}{d\frac{3}{2}k \alpha' \Delta}}\,.
\eeq
On the other hand, from  Eq.~\eqref{5.5b}  we have that the viscosity parameter $\z$ is such that\footnote{ 
Note that, for the sake of simplicity, we have chosen to represent the bulk viscosity in terms of the Hubble parameter. However, it should be emphasized that the reverse transformation is also achievable.}
\begin{equation}\label{zeta generic}
 \zeta= \frac{\tilde{Y}}{3}\le(\frac{d}{2}H(1-\om)-d H^3 \frac{\Delta \alpha'}{8} k (1-3 \om)\ri)\,.
\end{equation}

\subsection{Evaluating the stability in general case}
\label{SEC6.1}
We now want to compute the stability of the above solution. To this aim, we should compute the Jacobian of the system
$(\dot H (H, \dot \phi), \,\ddot \phi (H, \dot \phi) )$, evaluated at the SHG solution.

We begin by writing the functions $\mc{F}_\r, \mc{F}_\s, \mc{F}_p$ as
\bea\label{functions fab}
&&\mc{F}_\r= \mc{F}_\r(\fd,H)\,,\nonumber\\
&&\mc{F}_\s= \mc{F}_\s(\fdd,\Hd,\fd,H)\equiv \Hd\mc{F}_{\s,1}(\fd,H)+\fdd \mc{F}_{\s,2}(\fd,H)+\mc{F}_{\s,3}(\fd,H)\, ,\nonumber \\
&&\mc{F}_p= \mc{F}_p(\fdd,\Hd,\fd,H)\equiv \Hd\mc{F}_{p,1}(\fd,H)+\fdd \mc{F}_{p,2}(\fd,H)+ \mc{F}_{p,3}(\fd,H)\, ,
\eea
and the equations of motions of Eqs.~\eqref{EOM first order O(d,d)}  in the following form:
\bea \label{funct R}
&&\r= {\tilde{Y}\over 2} E_0(\fd,\Hd)\,,\nonumber\\
&&\s=-\tilde{Y} \le[\Hd E_{2,1}+\fdd E_{2,2}+ E_{2,3}   \ri]\,,\nonumber\\
&& p_{eff}={\tilde{Y}\over 2d} \le[\Hd E_{3,1}+\fdd E_{3,2}+ E_{3,3}   \ri]\,,  
\eea
where the values of $\mc{F}_{\a, n} $ and $E_{\a, n}$ are reported in  Appendix \ref{Ap.A}.

It is then simple to see that the first equation of Eqs.~\eqref{funct R} can be used as a constraint, such that the equations of motion can be written as
\begin{eqsplit}\label{system c1}
    \Hd E_{3,1} +\fdd E_{3,2} = C_1\,, \\
   \Hd E_{2,1} +\fdd E_{2,2} = C_2 \,, \\
\end{eqsplit}
where we use $\s=2p_{eff}$ and $p_{eff}=p-d H\zeta=\om\r-d H \zeta$, and we define
\begin{eqsplit}\label{eq for C1}
&C_1=-2 d^2 \zeta H \tilde{Y}^{-1}+\om d E_0-E_{3,3}\,,\\
&C_2=2 d\zeta H \tilde{Y}^{-1} -\om E_0-E_{2,3}\,.
\end{eqsplit}
The solutions for the system are straightforward 
\begin{equation}\label{Eq.5.13}
    \Hd =  \frac{C_1 E_{2,2}-C_2 E_{3,2}}{E_{3,1}E_{2,2}-E_{3,2}E_{2,1}}\equiv {r_1\over r}\,, \qquad \fdd=\frac{C_2 E_{3,1}-C_1 E_{2,1}}{E_{3,1}E_{2,2}-E_{3,2}E_{2,1}}\equiv {r_2\over r}\,,
\end{equation}
and the Jacobian is given by:
\beq
\mc{J}=\begin{pmatrix}
{\pa_H \Hd}\,,& \pa_\fd \Hd\\
\pa_H \fdd \,,& \pa_\fd \fdd
\end{pmatrix}=
{1\over r^2}\begin{pmatrix}
{r_1\pr r -r_1 r\pr}\,, & \dot{r}_1 r -r_1 \dot{r}\\
{r_2\pr r -r_2 r\pr}\,,&  \dot{r}_2 r -r_2 \dot{r}
\end{pmatrix}\,,
\eeq
here we use $\pa_H(\dots) \equiv(\dots)\pr$ and $\pa_\fd(\dots) \equiv\dot{(\dots)}$.

Evaluating the Jacobian at the fixed point, in order to have an attractor, we should have both eigenvalues negative, or equivalently, we should have
\begin{equation}\label{trace cond}
  \mathrm{Tr} \mathcal{J}|_{(H_*,\fd_*)}<0 \quad\, , \quad \mathrm{Det} \mathcal{J}|_{(H_*,\fd_*)}>0\,.  
\end{equation}
In order to solve such inequalities we can ignore the positive factor  $ \frac{1}{r^2} $ in the Jacobian.

\subsection{Numerical Results}
For graphical reasons, in the following
we have rescaled the Hubble constant and the viscosity parameter $ \zeta $ as $ H= \frac{h}{l_s} $ and 
$\zeta= \tilde \zeta \frac{ \tilde{Y}}{l_s}$, using the following values for the parameters: $ k=1$, $d=3$ and  $ 2 \pi  \alpha^{\prime} = l_s^2 $. 
We then impose the inequalities in Eqs.~\eqref{trace cond} for trace and determinant (for the explicit calculation see Eq.~\eqref{trace GMV} and Eq.~\eqref{det GMV} in appendix \ref{Ap.A}).  We made then the substitution $ \tilde \zeta= \tilde \zeta (h) $, in both the determinant and the trace and impose the constraint $ \tilde \zeta>0 $. Moreover, we impose the condition  $ C>0 $ that, using Eq.~\eqref{H generic}, implies for the GMV case $ h>\sqrt{\frac{8 \pi}{57}}$, and for the Meissner case $h>\sqrt{\frac{8 \pi}{3 }}$, where we have used the values of $\Delta$ in Eq.~\eqref{delta GMV} and Eq.~\eqref{delta Meisnner} respectively.

Those conditions should be solved in terms of $ w $ and $ h $, in such a way to obtain a region of stability in the plane $(\om, h)$. 
The results for both GMV and Meissner cases are showed in the Fig.~\ref{figura1} (left and right panel respectively). We illustrate both cases with and without a bulk viscosity (the orange region and the black line respectively).
In particular we find a large interval where the stability is possible for certain intervals of $\om$ and $h$.
For the GMV-case, as we can see from the left panel of Fig.~\ref{figura1}, the admitted values of $ \om $ are in the following interval  $ -1  \lesssim \om \lesssim 3 $ while $h$ should be such that  $ 0.72 \lesssim h \lesssim 1.14 $ .
On the other hand, the stability region for the Meissner case 
is such that we have the follows allowed interval  $ 0  \lesssim \om \lesssim 3.5 $ and    $  h  \gtrsim 3.5 $
(see right panel of Fig.~\ref{figura1}).

\begin{figure}[h!]
\begin{minipage}{0.5\textwidth}
\centering
\includegraphics[scale=0.9]{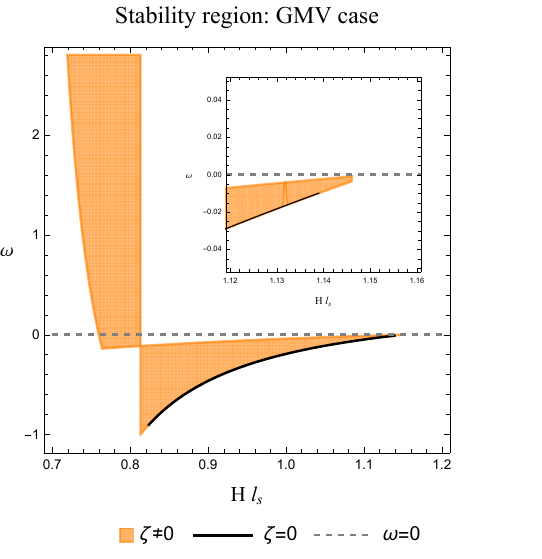}
\end{minipage}
\begin{minipage}{0.5\textwidth}
\centering
\includegraphics[scale=0.9 ]{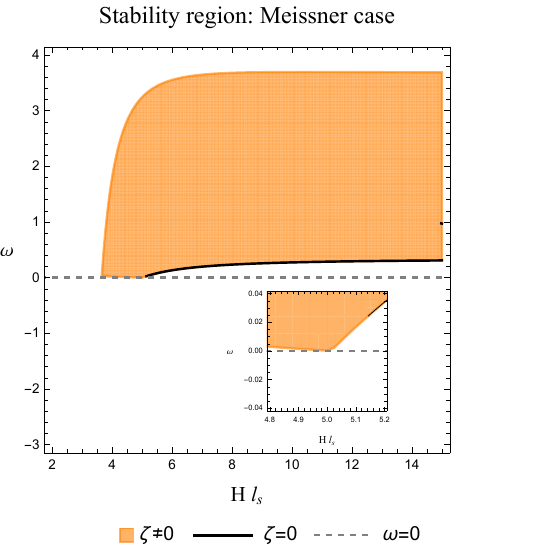}
\end{minipage}
\caption{We show the stability region in the plane $(\om, h)$, for the GMV case (on the left panel) and Meissner case (on the right panel). We plot the allowed region for $\z \neq 0$ (orange region) and for $\z=0$ 
(black line). }
\label{figura1}
\end{figure}

We find in both cases a large stability region. Even if the bulk viscosity $\zeta$ is negligible or zero we still have allowed values of the parameter space.  
Moreover, we note that, following the black line in the region plot that represents $ \zeta=0 $, we are in agreement with the results obtained in~\cite{Quintin:2018loc} for $\omega=0$. Indeed, we do not find any stability points for such a value of the equation of state.


\section{Discussion and Conclusions}
\label{sec.6}
In this manuscript we analysed a gas of string-hole as the final stage of the string phase in the pre-big bang scenario looking at the stability of the related solution. In particular, we extend the previous work~\cite{Quintin:2018loc}, firstly studying the string-hole gas (SHG) solution also in the case of a complete O(d,d)-invariant action at all orders in $\ap$. Furthermore, we considered the possibility to characterize the SHG as a fluid with a bulk viscosity term, without constraints on the equation of state. We analysed this possibility in the contexts of the low-energy action, first order in $\ap$ for the Gasperini-Maggiore-Veneziano and Meissner actions, and to all orders in $\ap$. 
We found that in the case of the low-energy action, with a local dilaton potential, the contribution of the viscosity does not change the conclusion obtained in \cite{Quintin:2018loc}. In all other cases the presence of viscosity allows us to have stable solutions. For the particular case of the the Hohm-Zwiebach action, we found that the solution can be reached requiring some condition for the general function $F(H)$, however only a sort of general stability is obtained. Indeed, the fixed point represent a ``center'' of the trajectory: we have circular orbits in the phase space, without touching the center of the circle.
We found also that, relaxing the constraint on the equation of state for the first order in $\ap$ (in~\cite{Quintin:2018loc} only the case $\omega=0$ was considered), stable solutions are allowed even in the case of negligible (or zero) bulk viscosity, as we can easily see in Fig.~\ref{figura1} for first order in $\ap$ corrected action.

Certainly it remains to study and describe with more accuracy the string-hole/black-hole formation and how the stringy nature of the black-holes manifest when they achieve the non-perturbative regime of string theory. Indeed, this manuscript is just a first step into this direction, with the attempt to take into account the non-ideal features of the SHG. 
As we have already mentioned, we are here considering a particular regime, where we are assuming a constant bulk viscosity. But a some point, when the SHG saturates the entropy bounds (\cite{Brustein:1997cv, Veneziano:1991ek}), such viscosity should go to zero. Therefore, we need to include a dynamical evolution of the parameters of the theory, to describe a complete evolution of the fluid.
Moreover, as already mentioned in the Introduction, a natural extension of the model is the inclusion of a non-zero shear viscosity starting from anisotropic initial conditions. 
Furthermore, it would be very interesting to study the dynamical evolution of the SHG within the recent ``hamiltonian'' formalism described in~\cite{Gasperini:2023tus, Conzinu:2023fth}. We postpone those analysis to future works.

\section*{Acknowledgements}
We are very thankful to Robert H. Brandenberger, Giuseppe Fanizza, Maurizio Gasperini and Gabriele Veneziano for useful discussions and feedback on the manuscript. 
We are supported in part by INFN under the program TAsP ({\it Theoretical Astroparticle Physics}). The work of G.M.
was partially supported by the research grant number 2022E2J4RK ``PANTHEON:
Perspectives in Astroparticle and Neutrino THEory with Old and New messengers''
under the program PRIN 2022 funded by the Italian Ministero dell’Universit\`a
e della Ricerca (MUR).

\appendix
\section{Details for the study at first order in $\ap$}
\label{Ap.A}

In this appendix we report the functions  introduced in 
Sec.~\ref{SEC6.1}. We use the following parametrization for the functions $\mc{F}_\a$
\beq\label{functions fab App}
\begin{split}
&\mc{F}_\r= \mc{F}_\r(\fd,H)\,,\\
&\mc{F}_\s= \mc{F}_\s(\fdd,\Hd,\fd,H)\equiv \Hd\mc{F}_{\s,1}(\fd,H)+\fdd \mc{F}_{\s,2}(\fd,H)+\mc{F}_{\s,3}(\fd,H)\, , \\
&\mc{F}_p= \mc{F}_p(\fdd,\Hd,\fd,H)\equiv \Hd\mc{F}_{p,1}(\fd,H)+\fdd \mc{F}_{p,2}(\fd,H)+ \mc{F}_{p,3}(\fd,H)\, .
\end{split}
\eeq
The function $E_0$ of Eqs.~\eqref{funct R}
is given by
\beq
E_0=\fd^2+d(d-1)H^2-2dH\dot \phi-{3\over 4} k\ap \mc{F}_\r\,,
\eeq
where the values of the functions $E_{\a,n}$ are
\begin{eqsplit}
    &E_{2,1}= 2d +{k \ap \over 4} \mc{F}_{\s,1}\,,\\
    &E_{2,2}= -2 +{k \ap \over 4} \mc{F}_{\s,2}\,,\\
    &E_{2,3}=\fd^2+d(d+1)H^2-2d H \fd +{k \ap \over 4} \mc{F}_{\s,3}\,,
\end{eqsplit}
and 
\begin{eqsplit}
    &E_{3,1}= -2d(d-1) +{k \ap \over 4} \mc{F}_{p,1}\,,\\
    &E_{3,2}=2d +{k \ap \over 4} \mc{F}_{p,2}\,,\\
    &E_{3,3}=-d^2(d-1)H^2 -d\fd^2+2d(d-1) H \fd +{k \ap \over 4} \mc{F}_{p,3}\,,\\
\end{eqsplit}
where, for simplicity, we omit the arguments of the functions $\mc{F}_\a$.

The parameters $ c_1 $ and $ c_3 $  which, as we will see shortly, appear in the functions $ \mc{F}_{\alpha} $ are given by
\bea
c_1&&=-\frac{d}{3}(d-1)(d-2)(d-3)\,,\nonumber\\
c_3&&=\frac{4d}{3}(d-1)(d-2)\,.
\eea

Moreover for the particular case of the SHG solution ($\dot\f=dH= \text{const.}$) we have the following relations between $ \mathcal{F}_{\rho} \, , \mathcal{F}_{\s}\, , \mathcal{F}_{p} $ . 
\begin{subequations}\label{delta}
    \begin{align}
    \mathcal{F}_{\rho}=&\mathcal{F}_{\sigma}= -d \Delta H^4\,,\\
    \mathcal{F}_p=& d^2 \Delta H^4=-d\mathcal{F}_{\rho}\, . 
    \end{align}
\end{subequations}

We now report the various functions and parameters for the particular cases under consideration: the Meissner and GMV actions.

\subsection{Meissner case}
The particular functions  $ \mc{F}_{\a,n} $ used in \eqref{functions fab App} are

\begin{equation}
\begin{split}
\mc{F}_\r(\fd,H)=&~c_1H^4+c_3H^3\dot \phi-2d(d-1)H^2\dot \phi^2+\frac{4}{3}dH\dot \phi^3-\frac{1}{3}\dot \phi^4\,,\\
\\
    \mc{F}_{\s,1}(\fd,H)=&~3c_3H^2-8d(d-1)H\dot \phi+4d\dot \phi^2 \, , \\
    \mc{F}_{\s,2}(\fd,H)=&~-4d(d-1)H^2+8dH\dot \phi-4\dot\phi^2\, , \\
    \mc{F}_{\s,3}(\fd,H)=&~(c_1+dc_3)H^4-4d^2(d-1)H^3\dot \phi+2d(3d-1)H^2\dot\phi^2-4dH\dot \phi^3+\dot \phi^4 \,, \\
    \\
    \mc{F}_{p,1}(\fd,H)=&~ 12c_1H^2+6c_3H\dot \phi-4d(d-1)\dot \phi^2 \, , \\
    \mc{F}_{p,2}(\fd,H)=&~ 3c_3H^2-8d(d-1)H\dot \phi+4d\dot \phi^2 \, , \\
    \mc{F}_{p,3}(\fd,H)=& ~3dc_1H^4-2(2c_1-dc_3)H^3\dot \phi-(3c_3+2d^2(d-1))H^2\dot \phi^2+4d(d-1)H\dot \phi^3-d\dot\phi^4\,,
\end{split}
\end{equation}

In this case, the quantity $ \Delta $ appearing in Eqs.~\eqref{delta}, and in the solutions \eqref{H generic} and \eqref{zeta generic} is given by 
$\Delta=d-2 $, such that
\begin{subequations}\label{delta Meisnner}
    \begin{align}
    \mathcal{F}_{\rho}=&\mathcal{F}_{\sigma}= -d(d-2) H^4\,,\\
    \mathcal{F}_p=& d^2 (d-2) H^4\, . 
    \end{align}
\end{subequations}

The explicit form of trace and determinant, in terms on the variables $h$ and $\omega$, and on the function $\tilde \zeta$,  which we have used in the inequalities in Eqs.~\eqref{trace cond} to study the stability of the SHG solution, are
\begin{equation}\label{trace Me}
\begin{split}
l_s r^2 ~\mathrm{Tr} \mathcal{J}|_{(H_*,\fd_*)}=&
\frac{1}{16 \pi ^4}(27 h (3 h^8 (51 w-52)-240 \pi  h^6 (2 w-3)-648 \pi  h^5\tilde \zeta + \\ & +16 \pi ^2 h^4 (9 w+35)+192 \pi ^2 h^3 \tilde\zeta +128 \pi ^3 h^2 (w-9)+384 \pi ^3 h \tilde\zeta +768 \pi ^4)
\, ,
\end{split}
\end{equation}
\newline
\begin{equation}\label{det Me}
\begin{split}
  l_s^2 r^4 ~\mathrm{Det} \mathcal{J}|_{(H_*,\fd_*)}=& 
-\frac{1}{1024 \pi ^8}243 h (21 h^4-24 \pi  h^2+16 \pi ^2) (189 h^{13} (21 w^2-286 w-19)+ \\ & -72 \pi  h^{11} (375 w^2-4818 w-713)-504 \pi  h^{10} (21 w+65)\tilde\zeta +\\& +48 \pi ^2 h^9 (1461 w^2-14814
   w-4451)+3168 \pi ^2 h^8 (19 w+67) \tilde\zeta +\\ &-384 \pi ^2 h^7 (\pi  (234 w^2-2368 w-746)+147 \tilde\zeta ^2)-384 \pi ^3 h^6 (393 w+473) \tilde\zeta+\\& +1536 \pi ^3 h^5 (\pi  (37 w^2-446 w-131)-3 \tilde\zeta
   ^2)+\\&+1536 \pi ^4 h^4 (85 w-3)\tilde \zeta -2048 \pi ^4 h^3 (2 \pi  (3 w^2-76 w-15)-21 \tilde\zeta ^2)+\\& -4096 \pi ^5 h^2 (9 w-25) \tilde\zeta -8192 \pi ^5 h (8 \pi  w+3 \tilde\zeta ^2)-65536 \pi ^6 \tilde\zeta
   ) 
\, .
\end{split}
\end{equation}
\subsection{Gasperini-Maggiore-Veneziano case}

The particular functions  $ \mc{F}_{\a,n} $ used in \eqref{functions fab App} are
\begin{equation}
\begin{split}
\mc{F}_\r(H,\dot\phi)\equiv&~ c_1{H}^4+c_3{H}^3\dot\phi-\dot\phi^4~,\\
    \\
    \mc{F}_{\s,1}(\fd,H)=&~3c_3H^2 \, , \\
    \mc{F}_{\s,2}(\fd,H)=&~ -12\dot\phi^2 \, , \\
    \mc{F}_{\s,3}(\fd,H)=&~(c_1+dc_3)H^4-4dH\dot \phi^3+3\dot \phi^4 \,, \\
    \\
    \mc{F}_{p,1}(\fd,H)=&~12c_1H^2 +6c_3H \dot \phi \, , \\
    \mc{F}_{p,2}(\fd,H)=& ~3c_3H^2 \, , \\
    \mc{F}_{p,3}(\fd,H)=& ~3dc_1H^4-2(2c_1-dc_3)H^3\dot \phi-3c_3H^2\dot \phi^2+d\dot \phi^4\, . 
\end{split}
\end{equation}
Here we have
 $\Delta=2d^2+d-2$, such that
\begin{subequations}\label{delta GMV}
    \begin{align}
    \mathcal{F}_{\rho}=&\mathcal{F}_{\sigma}= -d(2d^2+d-2) H^4\,,\\
    \mathcal{F}_p=& d^2 (2d^2+d-2) H^4\, . 
    \end{align}
\end{subequations}
 The trace and the determinant, functions of $ \om, h $ and of the function $ \tilde \zeta $ are the following:
\begin{equation}\label{trace GMV}
\begin{split}
l_s r^2 ~\mathrm{Tr} \mathcal{J}|_{(H_*,\fd_*)}= &-\frac{1}{16 \pi ^4}(27 h (57 h^8 (3783 w-2381) +\\& +48 \pi  h^6 (139 w+2725)-26928 \pi  h^5 \tilde\zeta -64 \pi ^2 h^4 (195 w+683)+\\&-22368 \pi ^2 h^3 \tilde\zeta +64 \pi ^3 h^2 (16 w+141)+\\&+3648 \pi ^3 h \tilde\zeta -768 \pi ^4))
\,,
\end{split}
\end{equation}

\begin{equation}\label{det GMV}
\begin{split}
      l_s^2 r^4 ~\mathrm{Det} \mathcal{J}|_{(H_*,\fd_*)}= &
-\frac{1}{16 \pi ^8}243 h (42 h^4-15 \pi  h^2+2 \pi ^2) (114912 h^{13} (399 w^2+995 w-152)+\\&-9 \pi  h^{11} (6150813 w^2+6771378 w-1634779)-306432 \pi  h^{10} (21 w-1) \tilde\zeta +\\&+12 \pi ^2 h^9 (1231395
   w^2+717546 w-315673)+288 \pi ^2 h^8 (54835 w-33881) \tilde\zeta+\\& -96 \pi ^2 h^7 (\pi  (9741 w^2-7952 w-3557)+\\&+18816 \tilde\zeta ^2)-96 \pi ^3 h^6 (21933 w-63143) \tilde\zeta -192 \pi ^3 h^5 (\pi  (641
   w^2+3143 w+74)+\\&+4332 \tilde\zeta ^2)-192 \pi ^4 h^4 (1157 w+6449) \zeta +512 \pi ^4 h^3 (\pi  (27 w^2+\\&+223 w-6)-291 \tilde\zeta ^2)+512 \pi ^5 h^2 (81 w+301) \tilde\zeta +\\&-1024 \pi ^5 h (8 \pi  w-27 \tilde\zeta
   ^2)-8192 \pi ^6 \tilde\zeta )
 \, .
\end{split}
\end{equation}

\bibliographystyle{JHEP}
\bibliography{Bibliography}

\end{document}